\documentclass[twocolumn,pre,aps,showpacs,showkeys,10pt]{revtex}
\usepackage{setspace}
\usepackage{graphicx}
\usepackage{amsfonts}
\begin{document}
\title{A drift formulation of Gresham's Law}
\author{Reginald D. Smith} 
\email{rsmith@bouchet-franklin.org}
\altaffiliation{PO Box 10051, Rochester, NY 14610}
\noaffiliation
\date{January 15, 2012}

\begin{abstract}
In this paper we analyze Gresham's Law, in particular, how the rate of inflow or outflow of currencies is affected by the demand elasticity of arbitrage and the difference in face value ratios inside and outside of a country under a bimetallic system. We find that these equations are very similar to those used to describe drift in systems of free charged particles. In addition, we look at how Gresham's Law would play out with multiple currencies and multiple countries under a variety of connecting topologies.
\end{abstract}

\maketitle

\section{An introduction to Gresham's Law}

While the general mechanism of economics, adjustment of supply and demand by feedback through pricing, is generally known and appreciated, few specific economic laws have wide awareness outside of the field like Gresham's Law. Gresham's Law is the frequently quoted (or mis-quoted) statement regarding the circulation of multiple species of legal tender commodity money in an economy where the overvaluing or undervaluing of either relative to face value per unit weight can cause a rapid shift in the availability of both where one can become scarce while the other becomes plentiful. In its most widely known formulation it states, ``bad money drives out good'', which is a shorthand to indicate the overvalued commodity currency becoming the dominant means of exchange in the money supply while the undervalued commodity currency becomes scarce in circulation as it is hoarded or exported.

At the outset though, we need to be more specific in defining terms. As remarked by Mundell \cite{mundell}, a more accurate (though not perfect) statement of Gresham's law is  ``Bad money drives out good if they exchange for the same price.'' The concept of overvaluing or undervaluing commodity money is due to the fact that a coin of certain weight and purity has a defined legal tender value by the issuer but the underlying commodity, typically gold or silver, has its own demand separate from the demand for money in the economy. Therefore, there can be a disconnect when the market value of the commodity of which the money is made differs significantly from the required legal tender value. When this occurs, one species becomes overvalued (``bad'') relative to the other currency while the other becomes undervalued (``good'') relative to the first. Since people will try to use the cheapest coin in a transaction given the same face value, the depreciated coin circulates while the other is hoarded or sold for a profit.

Gresham's Law has been long known, even before Sir Thomas Gresham to whom it is attributed. In fact, Gresham himself did not purport to have discovered this law but rather it was attributed to him by the British economist Henry Dunning Macleod in 1858 \cite{macleod} who believed Gresham had been the first to recognize this in a letter to Queen Elizabeth I advising her that the debasement of the coinage by her father Henry VIII had driven good coins out of circulation. The more exact stating of the phrase ``bad money drives out good'' was a paraphrasing of an anonymous 1696 pamphlet \emph{A Reply to the Defence of the Bank, setting forth the unreasonablesness of their slow payments}. Gresham's Law had been previously described even earlier by Nicole Oresme, a philosopher in the Middle Ages \cite{oresme} and Nicholas Copernicus, who needs no introduction, in an essay on coinage for Sigismund I, King of Poland \cite{copernicus}.

The operation of Gresham's Law has been viewed throughout history. In the Middle Ages there were three de facto commodity currencies: gold, silver, and copper and currency shortages due to debasement of one versus the other were common. There are also many recent historical examples. One is the ushering in of the global gold standard by the new nation Germany when in 1875 the Bank of Prussia declared it would only redeem banknotes in gold and not silver. While there were multiple causes for this, one was the decline in the silver price after large discoveries, particularly in the United States, which caused the silver coins to be overvalued relative to gold as the price of silver dropped. This caused an influx of silver into the country. The French led Latin Union and all other nations of Europe soon followed in demonetizing silver. Another example is in the United States where the de facto establishment of the gold standard in the United States was effected in 1834 by the 1834 Coinage Act which set the gold:silver price ratio at 16:1 versus the prevailing world market ratio of 15.6:1. This overvalued gold and caused an influx of gold  and the opposite effect on silver. Gresham's Law would be continuously invoked using this and other examples in the fierce debates on bimetallism later that century.

\section{A Mathematical Illustration of Gresham's Law}

In this paper, the key question we will address is how rapidly does commodity money outflow/inflow from a country when Gresham's Law comes into play. Evidence from times when Gresham's Law has operated in the past allude that the process can be quite rapid for even small differential price changes. The outflow in both quantity and total value can be shown to be quite large even in the most marginal circumstances.

\begin{figure}
\includegraphics[width=1.75in, height=1in]{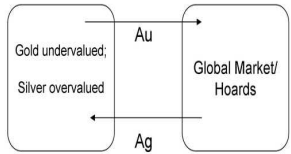}
\caption{Illustration of Gresham's law at work in a gold/silver bimetallic economy where gold is undervalued and silver is overvalued so the Au/Ag ratio is lower than that of the wider market.}
\label{goldsilver}
\end{figure}

In figure \ref{goldsilver} there are two countries (or a country and the wider market) which have open trade and capital flows allowing money of either species to circulate back and forth. If the ratio of the face values of the currencies in one country differ from the wider market, there will be net flows of each currency in each direction.  Here we will first illustrate the case where two species of commodity money, $m_1$ and $m_2$ circulate. Both have a global market price ratio of $R$, the ratio of the price of $m_1$ to $m_2$ and a local price in the domestic currency of $P_1$ and $P_2$. 

In addition, there are net demand functions for the commodity, $Q_1(P_1)$ and $Q_2(P_2)$ which describe the net demand curve for the currency over and above the demand for money when arbitrage opportunities emerge from mispricing. If $\frac{P_1}{P_2}=R$ then $Q_1=Q_2=0$. Further, assuming the overall money supply by value is unaffected, when Gresham's Law kicks in $P_1dQ_1=-P_2dQ_2$.

Given the above, we can thus state the demand function for $m_1$ alternatively as $Q_1(RP_2)=0$. The change of $Q_1$, $dQ_1$,  if there is a change in the face value of the coin from the market value of its metal can be given by

\begin{equation}
dQ_1 = -\frac{dQ_1}{dP_1}\Delta P_1
\end{equation}

The negative sign is indicative of the fact that $\frac{dQ_1}{dP_1}$ is negative for demand and $\Delta P_1$ greater than zero indicates an overvaluation which triggers inflows. This can be further expanded using the basic equation for elasticity of demand where the demand elasticity for $m_1$ is $E^1_d$

\begin{equation}
E^1_d = \frac{P_1}{Q_1}\frac{dQ_1}{dP_1}
\end{equation}

The equations can be combined to conclude

\begin{equation}
dQ_1 = -Q_1E^1_d \frac{\Delta P_1}{P_1}
\label{greshambasic}
\end{equation}

This shows that the differential quantity from equilibrium (in units of measurement such as tons) that flows out of the country is the product of the quantity demanded in the market, the elasticity of demand for the commodity arbitrage opportunity, and the \% increase of the face value over market value for the money. If you want to look at the outflow in terms of the unit currency we use $dM(1) = dQ_1P_1$ 

\begin{equation}
dM(1) = -Q_1E^1_d \Delta P_1
\label{greshambasic2}
\end{equation}

We use $dM(1)$ notation to differentiate from the money supply measure, $M_1$.

We can further reduce this by $\Delta P_1 = P_1 - RP_2$ to get

\begin{equation}
dQ_1 = -Q_1E^1_d \bigg(1-\frac{RP_2}{P_1}\bigg)
\end{equation}

\begin{equation}
dQ_1 = -Q_1E^1_d \bigg(1-\frac{R}{R_c}\bigg)
\label{greshamratio}
\end{equation}

where $R_c$ is the bimetallic ratio in the country that differs from the market.
Obviously if $R/R_c=1$ there is no net flow of currency. For the other species we have given $\Delta P_2 = P_2 - \frac{P_1}{R}$

\begin{equation}
dQ_2 = -Q_2E^2_d \bigg(1-\frac{R_c}{R}\bigg)
\end{equation}

Given the above, why is the outflow so sudden and huge in the case where face values and market values are not equal? First, the elasticity of demand $E_d$ is typically large (and negative) for the demand curves facing arbitrageurs \cite{arbitrage}. This combined with a likely high demand $Q_1$ by the market for the undervalued currency means $R_c \neq R$ will cause a huge flow under even a small differential between the market and legal tender ratios. This effect is only delayed by factors such as transaction and transport costs, historically estimated at about 1\% of face value for 19th century US coins. These factors show why bimetallic regimes are often unstable to even small changes in the face value or market price of the commodity used as currency. If $R$ is the Au/Ag ratio, when gold becomes overvalued, like the ratio being raised 16:1 by the 1834 Coinage Act instead of the 15.6:1 abroad, there is a positive net inflow for gold and the opposite for silver.

If there is a time constant, $\tau$ for the quantity of demand for a given currency, we can rewrite equation \ref{greshamratio} as

\begin{equation}
\frac{dQ_1}{dt} = -\frac{Q_1}{\tau}E^1_d \bigg(1-\frac{R}{R_c}\bigg)
\end{equation}

and show the quantity over time of a misvalued currency can be approximated by

\begin{equation}
Q_1(t) = Q_1(0)e^{-E^1_d \big(1-\frac{R}{R_c}\big)t/\tau}
\end{equation}

As a final note, the fact that $P_1dQ_1=-P_2dQ_2$ we have

\begin{equation}
 Q_1 E^1_d  \Delta P_1 = - Q_2 E^2_d  \Delta P_2
\end{equation}

\begin{equation}
 Q_1 E^1_d  \bigg(P_1 - RP_2\bigg) = - Q_2 E^2_d  \bigg(P_2 - \frac{P_1}{R}\bigg)
\end{equation}

\begin{equation}
Q_1 E^1_d = Q_2 E^2_d
\end{equation}

\begin{equation}
\frac{E^1_d}{E^2_d}=\frac{Q_2}{Q_1}
\end{equation}

\section{Gresham's Law and Drift}

For those familiar with physical processes of charge carriers such as those dealt with in plasma physics or materials science the equations above have a form that is qualitatively similar to charged particle drift under an external electric field. Whereas in those situations we have charge carriers of opposite sign, here we have currencies that are overvalued or undervalued. The classic charge density drift equation for a DC glow discharge plasma with a homogeneous electric field is \cite{plasmadrift}

\begin{equation}
J = qn\mu E
\end{equation}

Where $q$ is the unit charge of the charged particle, $e$ in the case of electrons, positrons, or protons, $n$ is the charged particle density, $\mu$ is the charge mobility, and $E$ is the electric field strength. A more comparable formulation to equations \ref{greshambasic} and \ref{greshambasic2} is

\begin{equation}
J = qn\mu \frac{\Delta V}{d}
\end{equation}

\begin{figure}

\includegraphics[width=3in, height=1.75in]{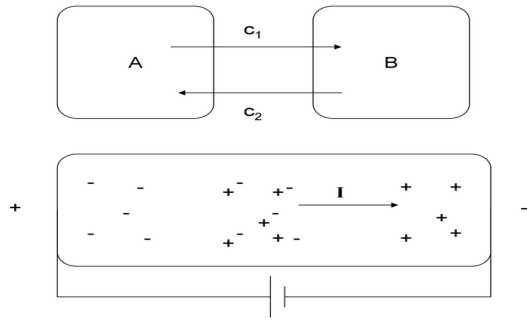}
\caption{A simple representation of the similarities between currency movements under Gresham's law and electric particle drift.}
\label{driftcompare}
\end{figure}

The variable $\Delta V$ is the voltage difference and $d$ is the distance between the electrodes. Roughly, the quantity demanded is comparable to $qn$, the demand elasticity to $\mu$, and the price differential to $\Delta V$.   This is illustrated in figure \ref{driftcompare}. 

\section{Expanded models of Gresham's Law}

For those interested in certain theoretical aspects of Gresham's Law beyond the real world typical case of bimetallic regimes, this section will address three cases of Gresham's Law under expanded scenarios of multiple currencies and countries. The number of currencies will be designated by $n$ and the number of countries designated by $m$. First, we will look at the situation with $m = 2$ countries and $n > 2$ currencies, then $n = 2$ currencies with $m > 2$ countries, and finally the complex case of $m > 2$ countries and $n > 2$ currencies. In all cases, all $n$ currencies are legal tender in all $m$ countries and have face values assigned by each country. In addition, the set of the values of the currencies are ordered sets so that each currency has a relatively higher or lower value to the same currencies across all countries. Finally, currencies will be labeled with numbers and countries with letters.

\subsection{Case of $m=2$, $n>2$}

In this case, we have $n$ currencies which will flow between the two countries in amounts corresponding to the number of available arbitrage opportunities. The value of any currency must be greater than zero so that the distribution of currency ratios for each currency pair in both countries is a closed set on $\mathbb{R}^+$. For each currency which has a face value that differs from the market value, it will have arbitrage opportunities against all of the other $n-1$ currencies as an outflow or inflow (or both). If a number of currencies, $C$, have different face values than market value in the other country, we have $\frac{nC}{2}$ possible arbitrage pairs with a maximum of ${n \choose 2}$ if everything is given a face value that differs from the market value abroad.

For a given currency, the net flow is given by

\begin{equation}
\Delta Q_i =  -\sum^n_{j=1} Q_{i,j} E_{d(i,j)}  \bigg(1-\frac{R_{i,j}}{R_{c(i,j)}}\bigg)
\end{equation}

Next we assume the demand function, and thus $E_d$ and $Q$ are identical across currency pairs for simplification.

\begin{equation}
\Delta Q_i =  -Q E_d  \sum^n_{j=1} \bigg(1-\frac{R_{i,j}}{R_{c(i,j)}}\bigg)
\label{greshamexpand1}
\end{equation}

From equation \ref{greshamexpand1} we can see that the summation can be simplified to $(n - \sum \frac{R_{i,j}}{R_{c(i,j)}})$ and thus the sum that determines the magnitude and direction of a currency's flow is

\begin{equation}
\frac{1}{n} \sum^n_{j=1}\frac{R_{i,j}}{R_{c(i,j)}}
\end{equation}

\begin{figure}
\includegraphics[width=1.75in, height=1in]{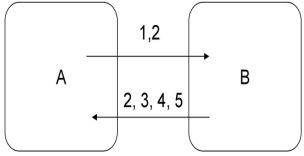}
\caption{The effect of overvaluing currency 2 in country $B$.}
\label{twocountry}
\end{figure}

If this quantity is less than 1 for a currency we have a net inflow, otherwise there will be a net outflow. From this it is obvious that the adjustment of the face value of just one of the currencies is enough to unsettle the entire equilibrium of currency flows between the two countries. For example (see figure \ref{twocountry}) imagine we have two countries, $A$ and $B$, with five currencies, 1 to 5. The value ratio of the currencies increases monotonically for a lower number over a higher number with $P_1/P_2 > 1$ and the highest ratio of all being $P_1/P_5$.  If country $B$ overvalues currency 2, but not so much that $P_1/P_2 < 1$, it will cause a drift in all the currencies. There will be a net inflow of currencies 1 and 2 since 2 is overvalued versus 3, 4, and 5 and 1 is overvalued versus 2. Currencies 3-5 will outflow since they are now relatively undervalued but 2 will also outflow since it is now undervalued with reference to currency 1. The net flow of currency 2 will depend on the relative volume of the arbitrage opportunity in each direction.

\subsection{Case of $m>2$, $n=2$}

In this case, we have multiple countries but only two currencies. At first glance, it would seem we could expand equations \ref{greshambasic} and \ref{greshamratio} to the case of summations across multiple countries to find the drift amount for currencies 1 and 2 in any given country $i$. However, here the actual behavior departs from idealistic physics characterizations and comparisons with charged particle behavior. With only two currencies and multiple countries there are multiple opportunities for arbitrage and given that the cost of transaction and transportation are similar, currencies will flow to the most profitable opportunity. So for currency 1 in a country $i$ the change to due arbitrage opportunities across all other countries indexed by $j$, we have

\begin{equation}
\Delta Q_1(i) = -Q_1E_d \max_j \bigg(1 - \frac{R(j)}{R_c(j)} \bigg)
\end{equation}

The implication from the maximization is that we minimize the quantity $\frac{R(j)}{R_c(j)}$. Further, the flow of currency 2 in the opposite direction is given by

\begin{equation}
\Delta Q_2(i) = -Q_2E_d \min_j \bigg(1 - \frac{R_c(j)}{R(j)} \bigg)
\end{equation}

\begin{figure}[h]
\includegraphics[width=3in, height=1.75in]{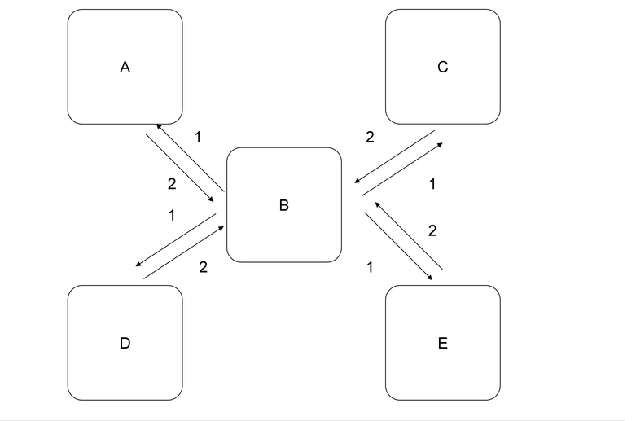}
\caption{The flows of currencies for $n=2$ and $m>2$.}
\label{lopsided}
\end{figure}

The minimization occurs given the flow is the opposite direction (negative) of currency 1. So here we must maximize $\frac{R_c(j)}{R(j)}$. Given this is the inverse of the previous minimization of $ \frac{R(j)}{R_c(j)}$ we find that the same country is the target of inflows and outflows of both currencies. If we consider the relative values of the currencies to be ordered sets with the same relative rank order for each currency in all countries, all countries will send the overvalued currency and receive the undervalued currency from the country with the most lopsided ratio. This is illustrated in figure \ref{lopsided} where currency 2 is most overvalued in country B.

\subsection{Case of $m>2$, $n>2$}

In the most complex case, it would seem nearly intractable given that at maximum disequilibrium you could theoretically have a ${n \choose 2}{m \choose 2}$ 
number of possible arbitrage trades. Granted, such a situation is likely never to arise since the proliferation of so many types of legal tender would defeat the purpose of money as a unit of exchange in the first place. In this situation, though, a  similar maximization strategy would appeal since in any country $i$, the currency $j$ would be sent to country $k$ where it is most overvalued relative to other currencies. In fact, if all $m$ countries are connected as in a complete graph, there is a country for each currency where it is most overvalued relative to its value in all others since the value of the currency must be positive. All other countries will send currency inflows of currency $j$ to this country given it provides the maximum return. 

It is possible for one country to have multiple currencies that are most overvalued, but each currency must have at least one country where it is most overvalued, multiple if some countries have the same relative value to the lowest currency. Again assuming a generic demand function for each currency to keep the quantity demanded and elasticity constant, 

\begin{equation}
\Delta Q_j(i) = -Q_j E_d \max_{k,l} \bigg(1 - \frac{R_{j,l}(k)}{R_{c(j,l)}(k)} \bigg)
\end{equation}

Again, we have to minimize $\frac{R_{j,l}(k)}{R_{c(j,l)}(k)}$ to maximize the entire argument. The lowest valued currency, with a default relative value of 1 in the ordered set, is thus undervalued in every country that has another currency most overvalued and flows out to all the other countries as the opposite outbound flow. Therefore, steady state will be a group of countries with only the currency(ies) that they most overvalue while the least value currency circulates throughout the other countries.

\begin{figure}
\includegraphics[width=1.75in, height=1.75in]{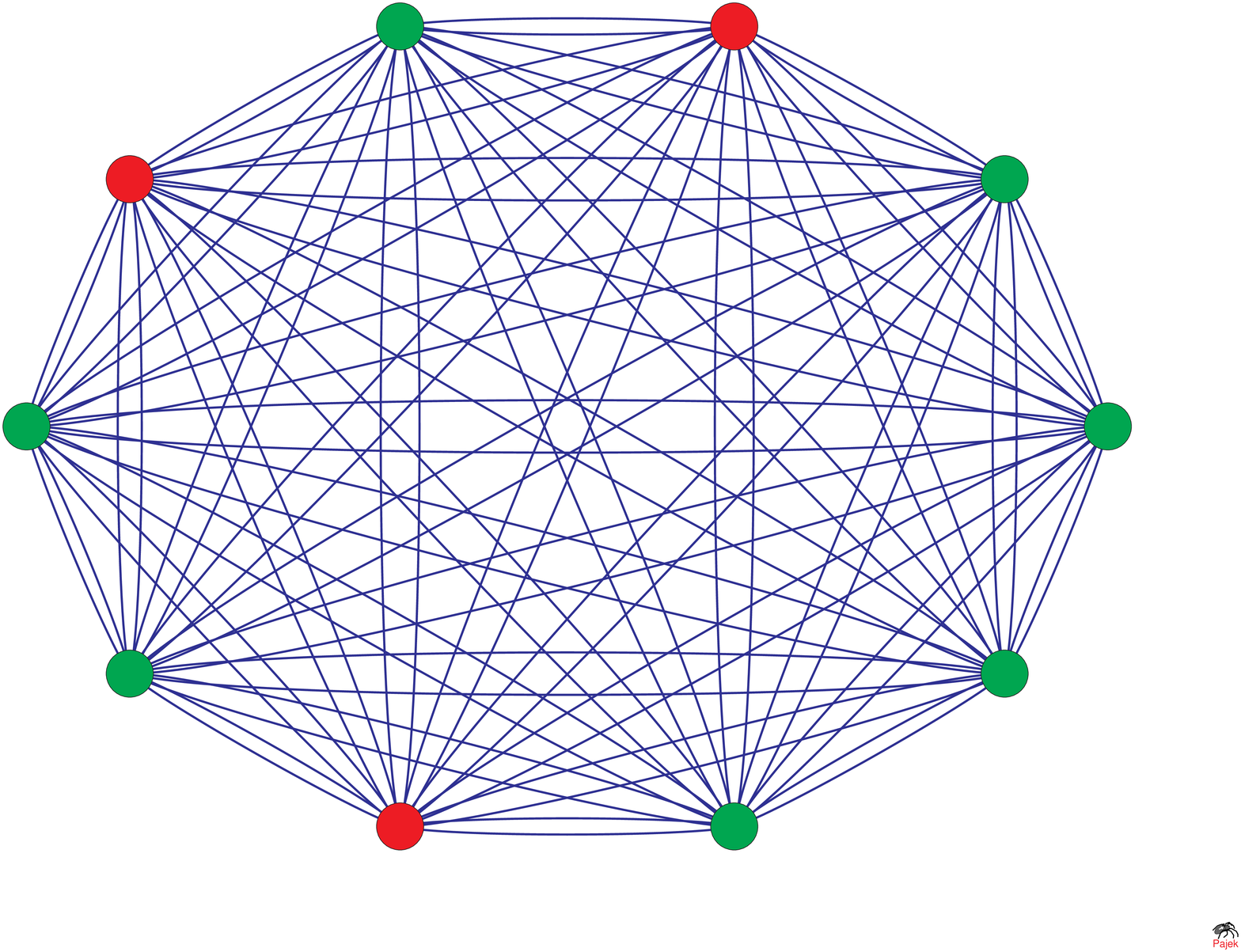}
\caption{Complete graph of $m$=10 with $n$=5. Red nodes show countries with inflows of overvalued currencies.}
\label{complete}
\end{figure}

\begin{figure}
\includegraphics[width=1.75in, height=1.75in]{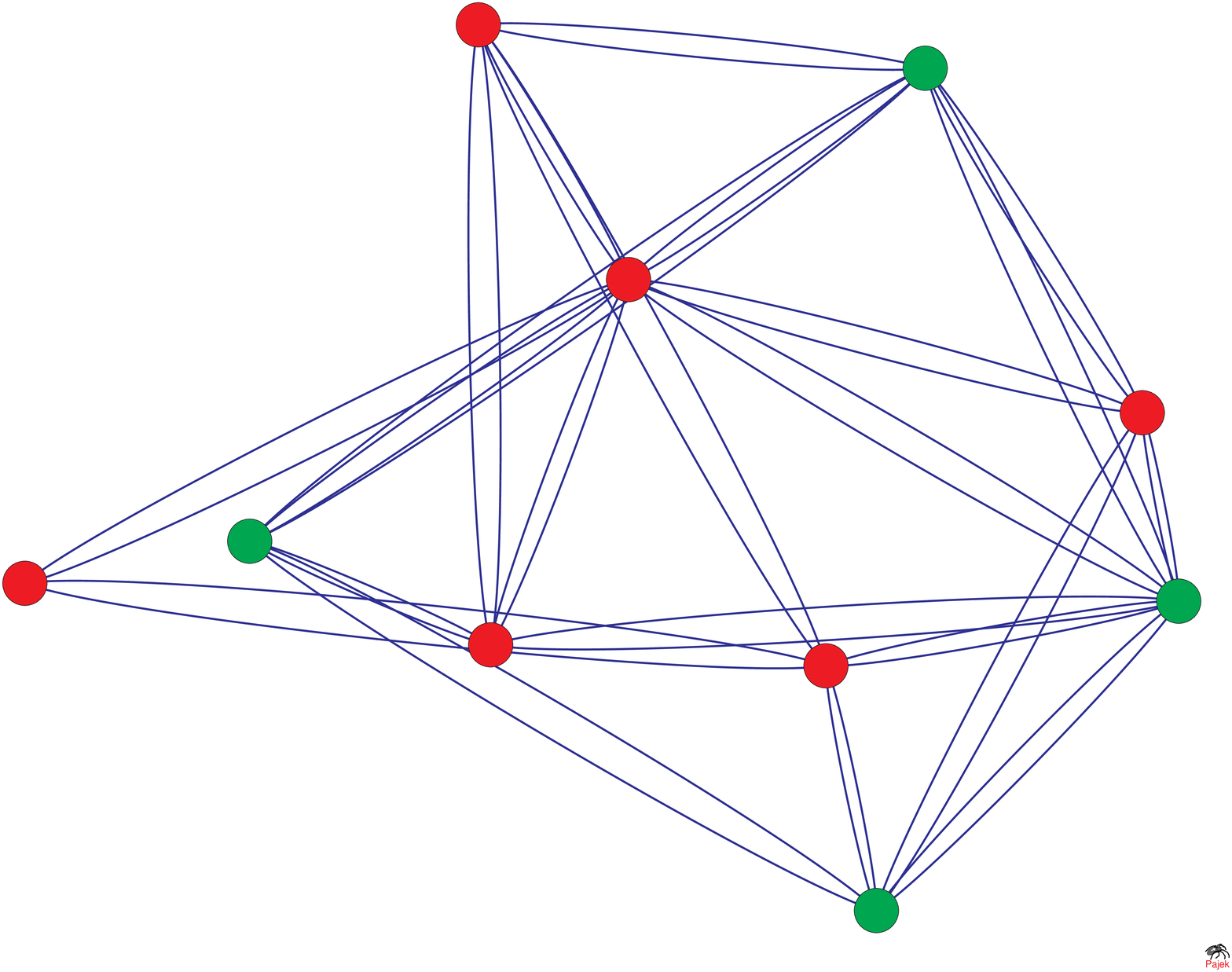}
\caption{Random graph of $m$=10, $n$=5 assuming $p$=0.48 for graph creation. Red nodes show countries with inflows of overvalued currencies.}
\label{random}
\end{figure}

\begin{figure}
\includegraphics[width=1.75in, height=1.75in]{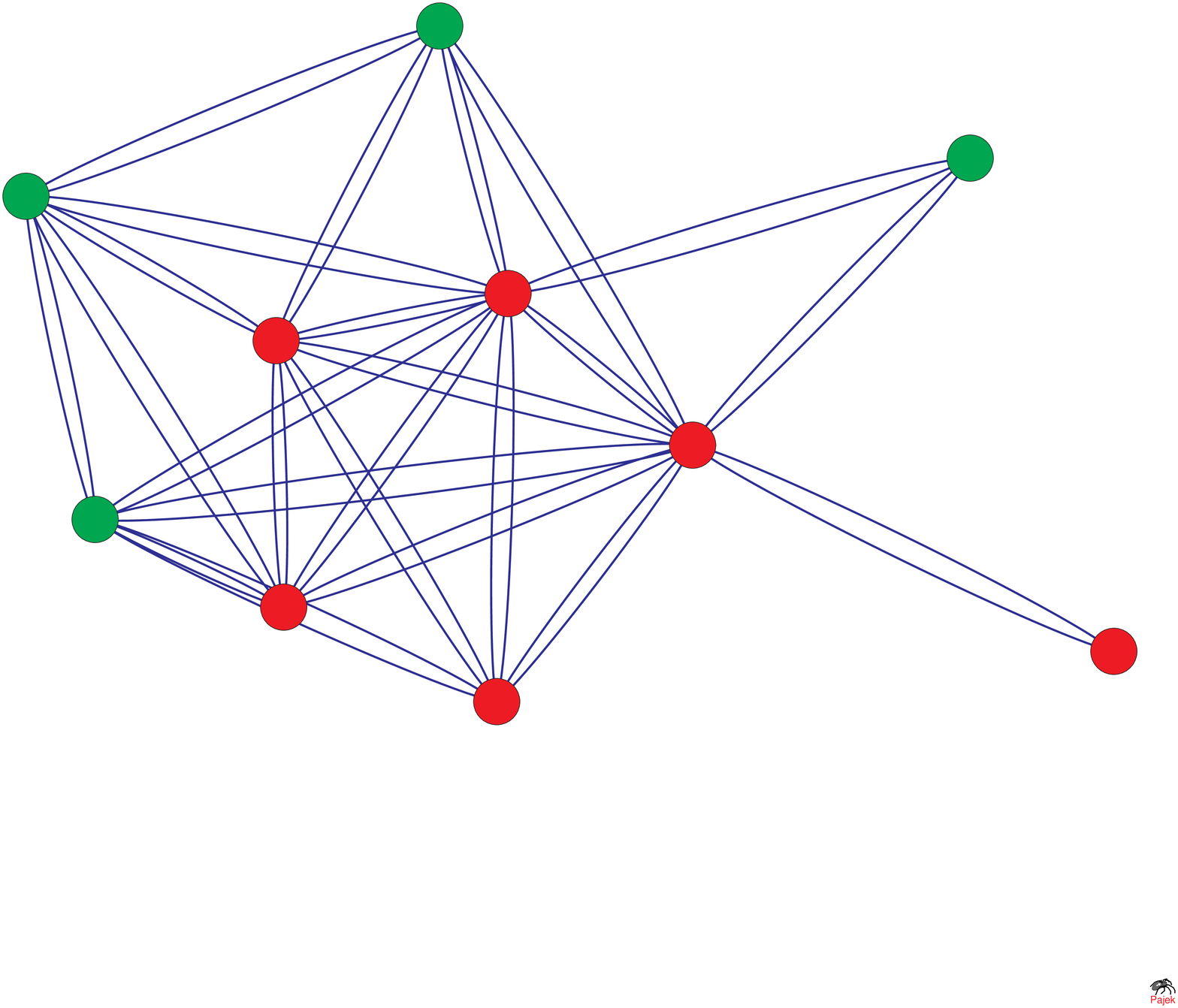}
\caption{Scale-free graph of $\langle k \rangle$=4.8. Red nodes show countries with inflows of overvalued currencies.}
\label{scalefree}
\end{figure}

Another possible situation under this model is where the topology is not one of a complete graph where every country (node) can have an inflow or outflow of currency to another. This could likely cause local maxima to arise given some nodes are not connected to the others where the currency is most overvalued. Indeed this is the case. In figures \ref{complete}, \ref{random}, and \ref{scalefree} we show the result of $m=10$, $n=5$ on the topologies of a complete graph, a random network, and a scale-free network.

In each graph, the red nodes indicate where one or more currencies are locally overvalued and thus have inflows from other nodes. Only in the complete graph are the red nodes indicative of global overvaluation since each node connects to every other. The latter two topologies have more nodes where currency is locally overvalued and have inflows despite not being a global maximum. Indeed, if one repeats the simulation 1000 times we find the results in the table \ref{compare} to show that there are more local maxima due to inhomogeneities in these stochastic graphs than the complete graph situation. Therefore, overvaluation and undervaluations of currency are more likely to cause local rather than global flows across all the countries.

\begin{table}
\centering
\small
\begin{tabular}{|c|c|c|c|}
\hline 
&Complete&Random&Scale-Free\\
\hline 
Mean&3.9&6.4&5.8\\
Std. Dev.&0.9&1.3&1.4\\
\hline 
\end{tabular}
\caption{The mean and standard deviation of the number of vertices with locally overvalued currencies by topology from a 1000 run simulation where $m=10$, $n=5$, and $\langle k \rangle = 4.8$ for the random and scale-free networks.}
\label{compare}
\end{table}

\section{Conclusion}

Gresham's Law may not rise to the level of a principle of physics and indeed even in economics it is often carefully handled. For example, it does not apply to paper money where you can often have dual circulations of different banknotes in an unofficial capacity without one driving the other out of circulation. This is often seen with dollars in many countries, especially in the Western hemisphere, Euros in parts of Eastern Europe not yet on the currency, and the Renminbi in Hong Kong and Southeast Asia. In these cases, only one currency is legal tender and while one may have more value in the sense of relative exchange rates of fiat money, the bill itself has no underlying value.

In this paper it has been demonstrated that Gresham's Law is a phenomenon quite similar to drift where the difference in valuation sets up pull similar to a potential that drives inflows and outflows of currency in a bimetallic regime. The rate of this flow is rapid due to its relation with the usually large elasticity of demand for arbitrage and is also proportional to the degree of imbalance between the ratios of the two currencies in the country and outside of it in the wider market. Free flow with negligible costs and delays between countries leads to a relatively few countries with the highest overvaluation receiving inflows of all but the lowest denominated currency while connections through more complicated types of networks lead to inhomogeneities and local minima for currency inflows and outflows. In the end, however, in almost all scenarios bimetallic or multimetallic currency regimes are extremely unstable to fluctuations in the face value that differ from market value. It is partially for this reason that they were not preferred standards in the latter half of the 19th century.

\end{document}